\documentclass[a4paper,pre,twocolumn,amsmath,amssymb,unsortedaddress,showpacs]{revtex4-1}
\usepackage{graphicx}  
\usepackage{dcolumn}
\usepackage{bm}
\newcommand{\be}{\begin{equation}}
\newcommand{\ee}{\end{equation}}

\begin{document}

\title{Soliton complexity  in the
damped-driven nonlinear Schr\"odinger equation: stationary, periodic, quasiperiodic complexes}

 \author{I. V.  Barashenkov}
\email{Email: igor.barashenkov@uct.ac.za}
\affiliation{Department of  Mathematics,
University of Cape
Town, Rondebosch 7701;}
 \affiliation{National Institute for Theoretical Physics, Stellenbosch, South Africa}
  \affiliation{Joint Institute for Nuclear Research, Dubna, 
 Russia}
 
\author{E. V.  Zemlyanaya}
\email{Email: elena@jinr.ru}
\affiliation{Joint Institute for Nuclear Research, Dubna, 141980 Russia}

\date{\today}

\begin{abstract}
Stationary and oscillatory bound states, or complexes,  of the damped-driven 
 solitons are numerically path-followed  in the parameter space.
 We compile 
a chart of the two-soliton
attractors, complementing the one-soliton attractor chart.
\end{abstract}

\pacs{05.45.Yv}

\maketitle

%
\section{Introduction}
\label{Intro}
This paper continues the study of localised time-periodic solutions of the parametrically driven
damped  nonlinear Schr\"odinger equation,
\be 
i \psi_t + \psi_{xx} + 2 |\psi|^2 \psi - \psi = h \psi^* - i
\gamma \psi.
\label{NLS} 
\ee
(Here $h,\gamma>0$).
Equation \eqref{NLS}
is an archetypal  equation for small and slowly-varying amplitudes of 
waves and patterns in 
spatially-distributed parametrically driven systems. It was employed to model 
intrinsic localised modes in coupled microelectromechanical and nanoelectromechanical 
resonators \cite{MEMS}, solitons in dual-core nonlinear optical fibers \cite{DCF}
and
dissipative structures in  optical parametric oscillators \cite{OPO}.
The discrete version of \eqref{NLS} was studied as a prototype for the energy
localisation in nonlinear lattices \cite{DB}.
(More contexts are listed in \cite{BZvH}.)

In the previous publication \cite{BZvH}, its authors proposed to obtain the time-periodic solitons
as solutions of the two-dimensional boundary-value problem
with  the boundary conditions
\be
\psi(x,t) \to 0 \quad \text{as} \ x \to \pm \infty;
\quad
\psi(x,t+T)=\psi(x,t).
\label{BC} \ee
In the present paper, we apply this approach to the analysis of {\it complexes\/} of 
solitons.

Complexes (also known as molecules) are stationary or oscillatory associations of two or more solitons;
they can be stable or unstable. Stable solitonic complexes, or bound states, were detected
in a variety of soliton-bearing partial differential equations     \cite{comp,Mal_Par,tails,embed,BZ1,water,BW,long_co,Baer,BZ2}.
One mechanism of complex formation
   is the trapping of the soliton in
a potential well formed by the undulating tail of its partner 
\cite{Mal_Par, tails}. 
This mechanism is not accessible \cite{Mal_Par} to the 
parametrically driven damped solitons though,
 as their tails are
  decaying  {\it monotonically\/}.
  The exchange of resonant radiation can also serve as a binding formula
  in nondissipative systems
  \cite{Mal_Par,embed}, but in the damped-driven equation \eqref{NLS} the radiation is nonresonant.
  A different mechanism was shown to operate here, which relies on
  the phase-stimulated growth or decay of the soliton's mass \cite{BZ1}. 

Bound states serve as long-term attractors in situations 
where there is more than one soliton present in the initial condition. 
For example, two like-polarity surface
solitons in a vertically-driven water tank attract each other and form a stable bound state \cite{water}.
Unstable complexes do not have the same experimental visibility
and can appear only as transients in numerical simulations.
However,  unstable complexes have a mathematical role to play:
they work as the phase-space organisers \cite{BW}.

The formation of complexes with an increasing number of elementary
constituents \cite{long_co} gives rise to a higher degree of {\it spatial\/} complexity in the system, in the
same way as the binding of  shorter molecules into longer ones
produces chemical compounds with increasingly complex properties.
 Previous analyses  were confined to stationary \cite{BZ1} and steadily moving \cite{Baer,BZ2} associations 
 of the parametrically driven solitons.
In the present paper we extend these studies to 
time-periodic complexes, thereby increasing the {\it temporal\/} complexity of the localised
structures.

We consider time-periodic complexes as ``stationary" solutions of Eq.\eqref{NLS} 
on a two-dimensional 
domain $-\infty <x< \infty$,  $0 \leq t \leq T$.
This allows to determine both stable and unstable complexes. 
Solutions of the boundary-value problem \eqref{NLS}, \eqref{BC} are 
path-followed in the parameter space  --- in the same way as free-standing periodic solitons were 
continued in the previous publication \cite{BZvH}.

An outline of this  paper is as follows. In the next section we describe
bifurcations of the {\it static\/} two-soliton complexes.
Of particular importance here are the Hopf bifurcations; these give birth
to time-periodic solutions.
We establish that the values of the damping coefficient are divided into two ranges. Namely, 
for $\gamma$ larger than a certain threshold, the complex suffers one or more Hopf bifurcations as $h$ is varied.
Below the threshold $\gamma$, no Hopf bifurcations occur.

 In section \ref{Num2},
the Hopf-bifurcation points of the stationary complexes are exploited
as the starting points of the $T(h)$ curves for the 
 time-periodic complexes. These curves are traced as we continue the 
 periodic bound states  in 
a parameter. Depending on the number of the Hopf bifurcations suffered by the 
static complex, we have one, two or more branches of the periodic solutions
emanating out of it. 
Complexes resulting from different Hopf bifurcations follow different 
transformation routes.

In the concluding section (section \ref{DC}) the results on stationary and time-periodic complexes are summarised
in the form of a two-soliton attractor chart.
Included in this chart are also some {\it quasiperiodic\/}
attractors.

\section{Stationary two-soliton complexes}
\label{CP}
The two free-standing stationary soliton solutions to Eq.\eqref{NLS} 
are distinguished by the subscripts $+$ and $-$:
\begin{subequations}
\label{sol}
\begin{equation*}
\psi_\pm (x)= A_\pm  \exp (- i \theta_\pm) \,  {\rm sech\/} (A_\pm x), 
\label{psi}
\end{equation*}
where
\begin{equation*}
 A_\pm = \sqrt{1 \pm  \sqrt{h^2-\gamma^2}}, \quad
 \theta_+ = \frac12 \arcsin \frac{\gamma}{h},
\label{par}
\end{equation*}
\end{subequations}
and $\theta_- = \pi/2 - \theta_+$. The $\psi_-$ soliton is unstable for all $h$ and $\gamma$.
The soliton $\psi_+$  is stable when the difference $h-\gamma$ is small but loses its stability
to a time-periodic soliton when $h$ exceeds a certain limit $h_{\rm Hopf}(\gamma)$.

The solitons $\psi_+$ and $\psi_-$ can form a variety 
of bound states, or complexes \cite{BZ1,BZ2,Baer}. (For example, in the previous paper \cite{BZvH} we 
mentioned a complex $\psi_{(-+-)}$, that is, a symmetric stationary association
of two solitons $\psi_-$ and one $\psi_+$.)
All complexes involving the $\psi_-$ solutions are, expectably,  unstable;
however two $\psi_+$ solitons can form a stable bound state \cite{BZ1}.

Previously,  the two-soliton complex $\psi_{(++)}$
was  known to exist only for sufficiently
large values of damping \cite{BZ1,BZ2}.  
We have now established that this  complex exists for all $\gamma \geq
1.5 \times 10^{-8}$. Its domain of existence on the
$(\gamma,h)$-plane is not bounded from above
except that
 for $h$ greater than
  \[
 h_{\rm cont}=\sqrt{1+\gamma^2},
 \]
the complex  is unstable to the continuous-spectrum
perturbations (as any
other solution decaying to zero at the infinities).
 Reducing
$h$ from $h_{\rm cont}$ for the fixed $\gamma$, 
we obtain one of two possible types of
bifurcation diagrams on the $(h,E)$ plane,
where the energy $E$ is defined by
\begin{equation}
{ E}=\int_{-\infty}^{\infty}
\left[  |\psi_x|^2 + |\psi|^2 - |\psi|^4 + h \frac
{\psi^2+ {\psi^*}^2}{2} \right] dx.
\label{E}
\end{equation}
(The energy  is {\it not\/} an integral of
motion when $\gamma \neq 0$;
however,  $E$ is obviously a constant for time-independent
solutions and can be used as a physically meaningful
bifurcation measure.)

\begin{figure}
\includegraphics[width =\linewidth]{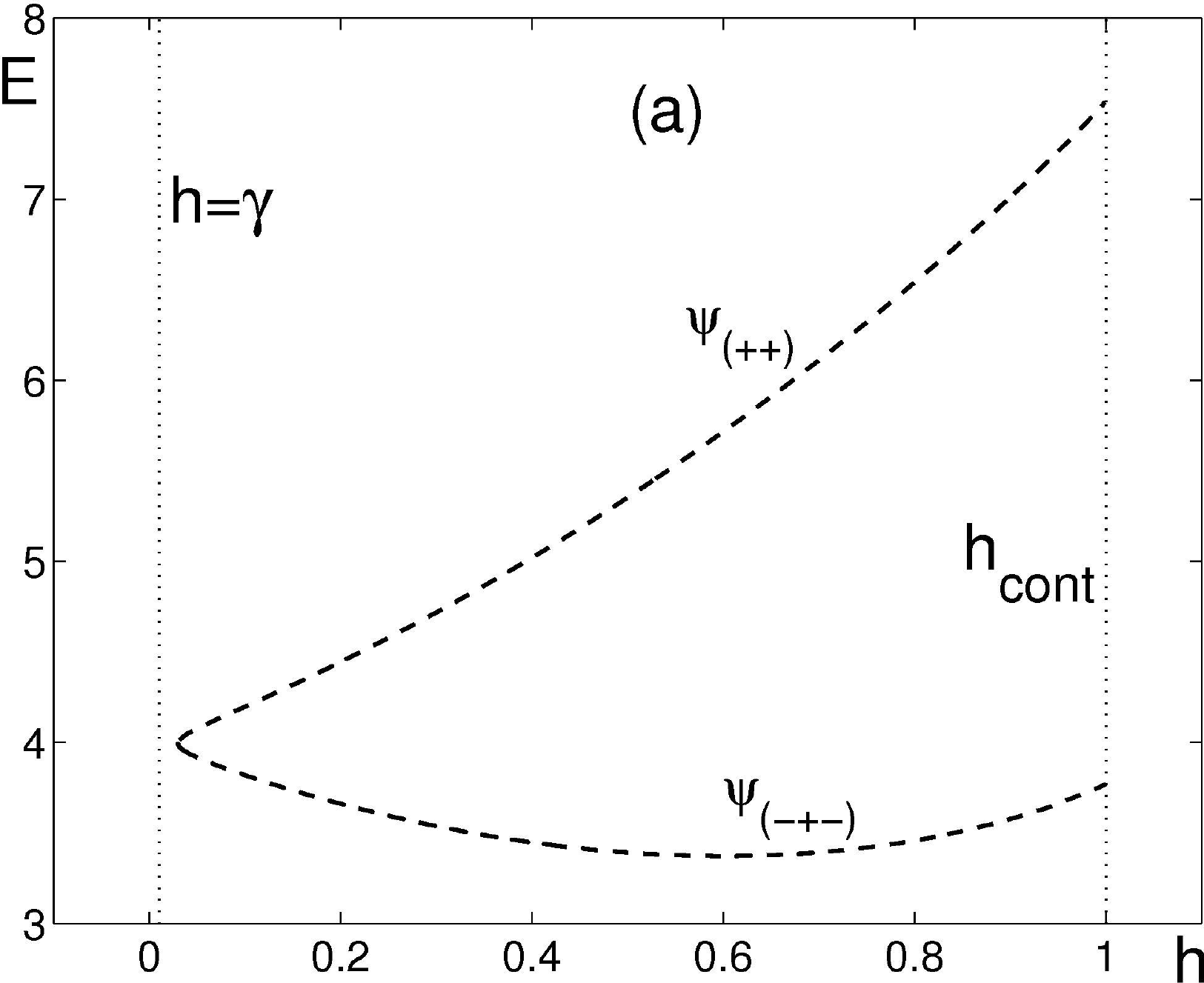} 

\vspace*{2mm}
\includegraphics[width = \linewidth]{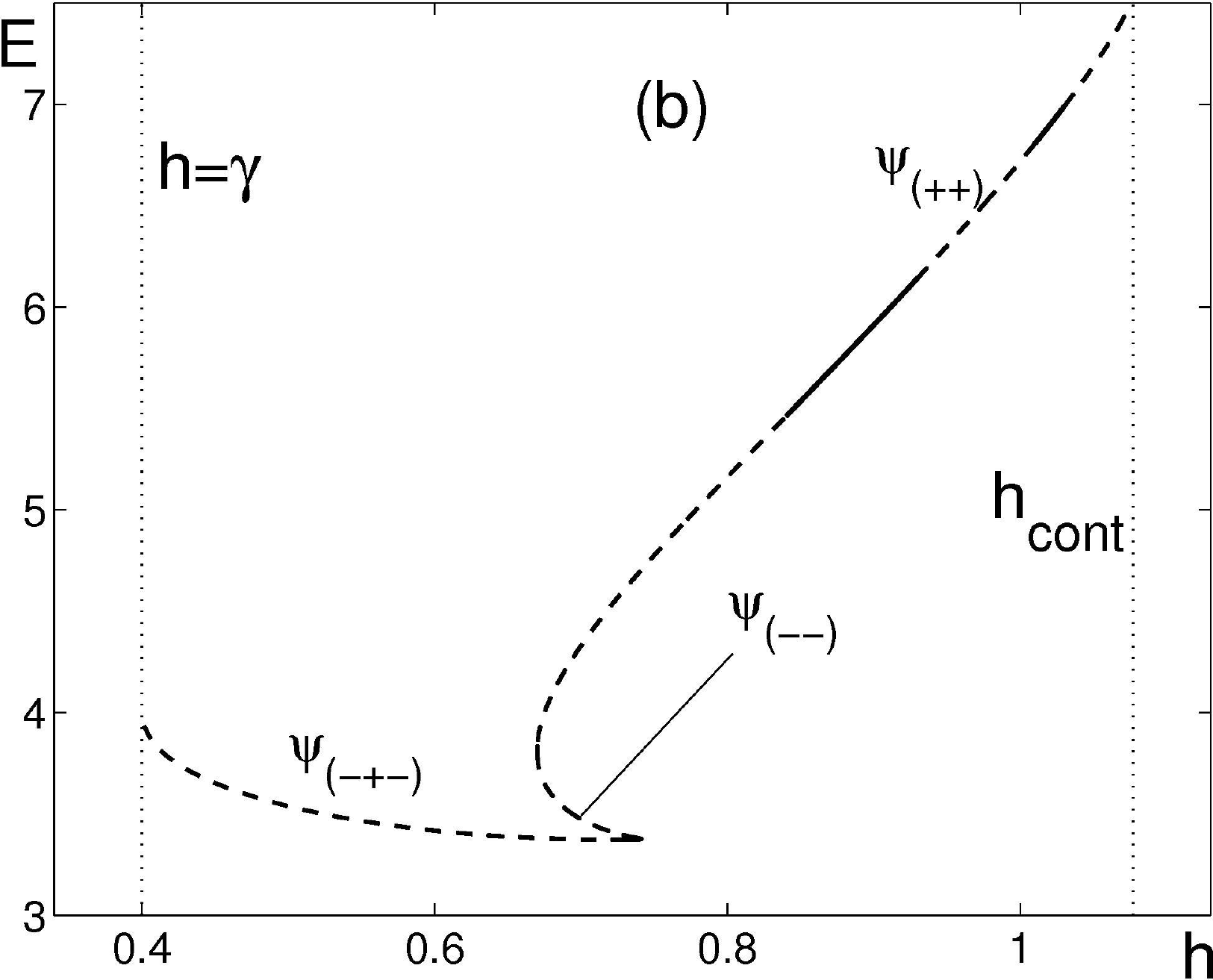}
\caption{\label{stat_complexes} Energy of the stationary 
two-soliton complex and
stationary multisoliton solutions obtained from this complex by
continuation in $h$ for the fixed $\gamma$. (a) $\gamma=0.01$; (b)
$\gamma=0.4$. Solid curves show stable and the dashed ones
unstable solutions. Note two  intervals of stability of the 
$\psi_{(++)}$ complex in (b).
}
\end{figure}	

\begin{center}
\begin{figure}[t]
\includegraphics[width = \linewidth]{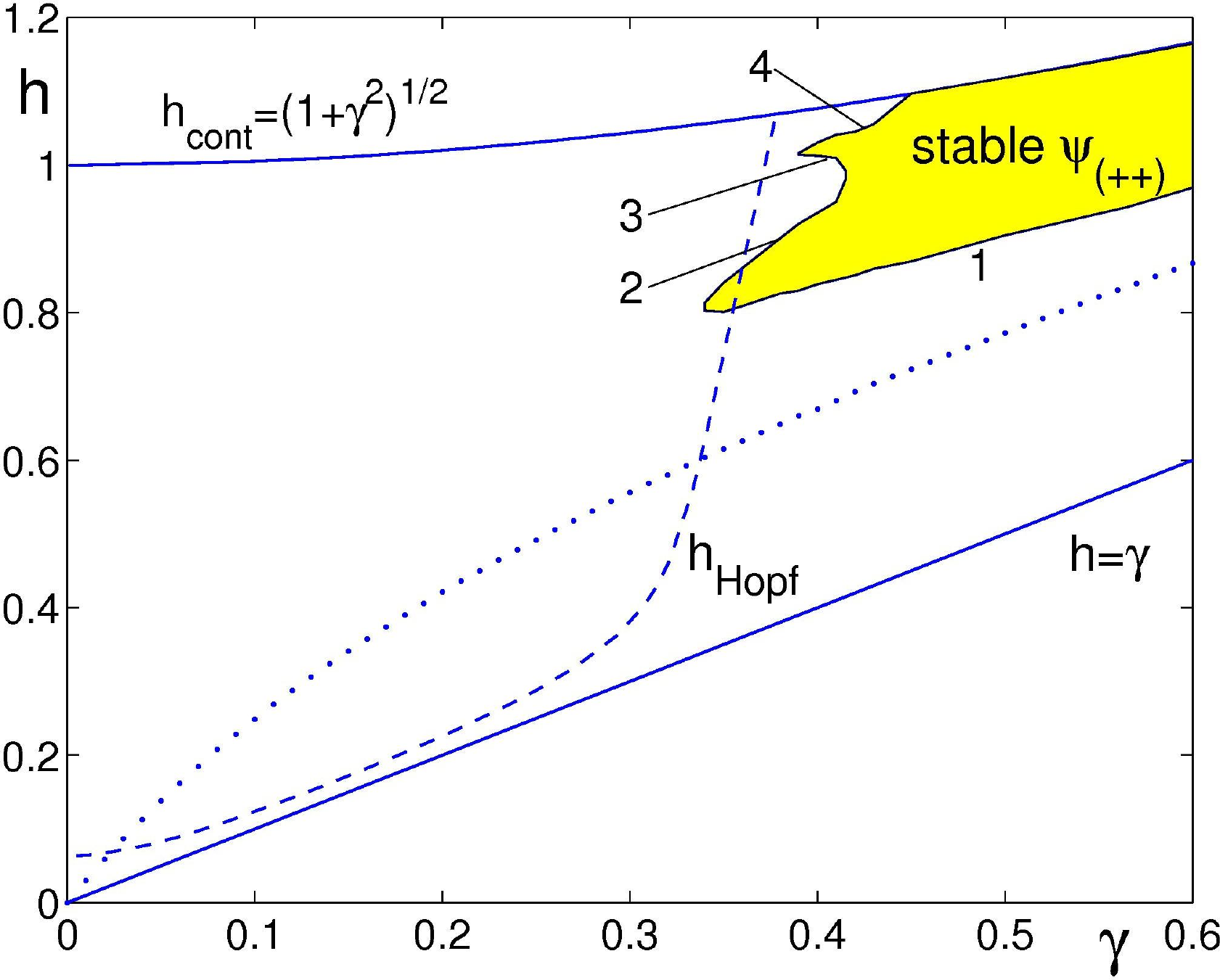}
\caption{(Color online)  The existence and stability domains of the  stationary
two-soliton complex $\psi_{(++)}$. 
The region of existence of the complex is  bounded from below by the dotted line; in the upwards direction
it extends  beyond
the value $h_{\rm cont}$, without bound. The complex is stable only in a small 
part of this region [tinted yellow (light gray)].  The stability domain
is bounded by the curve $h=h_{\rm cont}(\gamma)$ on the top
and by the lines of four Hopf bifurcations (1, 2, 3, and 4) on other sides. 
The dashed curve is the line of the 
Hopf bifurcation of the single soliton $\psi_+$. (The soliton is unstable {\it above\/} this line.)
\label{chart}} 
\end{figure}
\end{center}

The diagram of the first type [Fig.\ref{stat_complexes}(a)] arises
 when $h$ is decreased for a fixed {\it small\/} $\gamma$
($\gamma \leq 0.292$). In this case, there is only one turning
point, $h=h_{\rm sn}$, with $h_{\rm sn}=h_{\rm sn}(\gamma)  > \gamma$. 
[For
the parameter value $\gamma=0.01$ which we 
used to create Fig.\ref{stat_complexes}(a), 
$h_{\rm sn}=0.02972$; for $\gamma=0.1$, the turning point
is at $h_{\rm sn}=0.25$, and  for $\gamma=0.25$, $h_{\rm sn}=0.49$.] 
As $h$ approaches $h_{\rm sn}$ 
along the top branch, the two-soliton solution $\psi_{(++)}$  
develops a third hump halfway between the two humps that are
already there,
 with the distance between the lateral humps
remaining unaffected by this development. The complex obtained by the
continuation of this solution to the bottom branch can be identified
as a three-soliton bound state $\psi_{(-+-)}$.
 As we continue away from
$h_{\rm sn}$ along the bottom branch in
Fig.\ref{stat_complexes}(a), the $\psi_-$ solitons bound in this
complex (the two side solitons) diverge to the infinities on the
$x$ axis. 

All branches in the diagram of the first type consist of unstable solutions. 
(Our approach to the stability analysis of stationary solutions has been outlined in \cite{BZvH}.)

A somewhat different diagram arises for larger values of $\gamma$
($\gamma > 0.292$), see Fig.\ref{stat_complexes}(b). This
bifurcation diagram has been described in \cite{BZ1} for a particular
 $\gamma$ ($\gamma =0.565$); here, we reproduce
it for a different value of the damping coefficient. 
Reducing $h$ from $h_{\rm cont}$ for the fixed
$\gamma$, the  branch resulting from the two-soliton solution
$\psi_{(++)}$ develops two turning points instead of one. As we pass
the first turning point, the $\psi_{(++)}$ complex transforms into
the $\psi_{(--)}$ solution. Moving away from the point along
the bottom branch, the $\psi_{(--)}$ complex acquires a third
hump. This branch does not continue all the way to $h_{\rm cont}$ but turns
again, into a branch with even a lower energy. On this branch, the
three-hump solution can be identified as $\psi_{(-+-)}$. The
lowest branch continues  to the point $h=\gamma$. This point
defines the lower boundary of the domain of existence of the
stationary complexes which result from the path-following of the two-soliton solution $\psi_{(++)}$.
 As we approach the point $h=\gamma$, the distance
between  the two side solitons in the $\psi_{(-+-)}$ complex tends
to infinity.

For $\gamma$ just above  $0.292$, 
all solution branches in the diagram of the second type are unstable.
However, as $\gamma$ exceeds $0.34$, a stability window opens 
in the $\psi_{(++)}$ branch. 
The existence and stability domains of the two-soliton complexes on the $(\gamma,h)$ plane 
are shown in Fig.\ref{chart}.

We were not able to obtain a symmetric two- or
three-soliton complex for $\gamma=0$. If we 
fix $h$ and continue in $\gamma$
towards $\gamma=0$, the separation distance
between the solitons in the complex grows without
bounds; hence we conjecture that symmetric  multisoliton complexes
do not exist for $\gamma=0$. (There are {\it nonsymmetric\/}
complexes with $\gamma=0$ though; see \cite{Baer}.)

The shape of the $E(h)$ curve corresponding to  $\gamma=0.4$
[Fig.\ref{stat_complexes}(b)] looks similar to that of the $E(h)$ curve
for $\gamma=0.565$ \cite{BZ1}. The main difference
between the diagrams pertaining to these two values of $\gamma$ 
is that when $\gamma=0.565$, the stability region of the
two-soliton solution is seamless, i.e. does not have instability
gaps in it, whereas in the $\gamma=0.4$ case, the stability region
consists of two segments of the curve separated by an interval of
instability.
This difference is reflected by the shape of the stability
domain on the $(\gamma,h)$-plane
(note the  ``instability bay" on the north-west coast of  the stable region in Fig.\ref{chart}).

Each point of the ``coastline" of the stability ``continent"  in Fig.\ref{chart}  corresponds to a 
Hopf  bifurcation of the stationary complex
[except for points
along the curve $h=h_\textrm{cont}(\gamma)$]. 
The ``coastline"   consists of four segments
(marked 1, 2, 3, and 4 in Fig.\ref{chart}).
 Continuing  in $h$ along a vertical line $\gamma=const$
one crosses one, two or four of these;
accordingly, for a given $\gamma$, the 
complex may undergo one, two or four Hopf bifurcations.

\section{Time-periodic complexes}
\label{Num2}

The first segment (marked 1 in Fig.\ref{chart}) is defined as 
the ``south coast" of the tinted continent. It extends from $\gamma=0.34$
to larger $\gamma$
without a visible bound  --- presumably all the way to $\gamma=\infty$.
The line of the second  Hopf bifurcation (marked 2)
represents the ``north coast of the southern peninsula" in Fig.\ref{chart}; it 
is bounded by $\gamma=0.34$ on the left and $\gamma=0.413$ on the right.
The ``south coast of the northern peninsula" corresponds to the third 
Hopf bifurcation (marked 3); this extends from $\gamma=0.39$ to $\gamma=0.413$.
Finally, the top, fourth, Hopf bifurcation arises for $\gamma$ between 
$0.39$ and $0.445$ (marked 4). When $\gamma$ is greater than $0.445$, the complex 
undergoes just one Hopf bifurcation as $h$ varies (the one marked 1). 

\subsection{The first Hopf bifurcation}

In this subsection, we path-follow time-periodic complexes
born in the lowest Hopf bifurcation 
(i.e. detaching  from the south coast of the tinted ``continent"
in Fig.\ref{chart}). 
We take $\gamma=0.565$
as a representative value of damping  in the region where 
the stationary two-soliton complex undergoes only one Hopf bifurcation
and  $\gamma=0.35$ and $0.38$ in the 
region where there is more than 
one Hopf point. 

\begin{figure}[t]
\includegraphics[width = \linewidth]{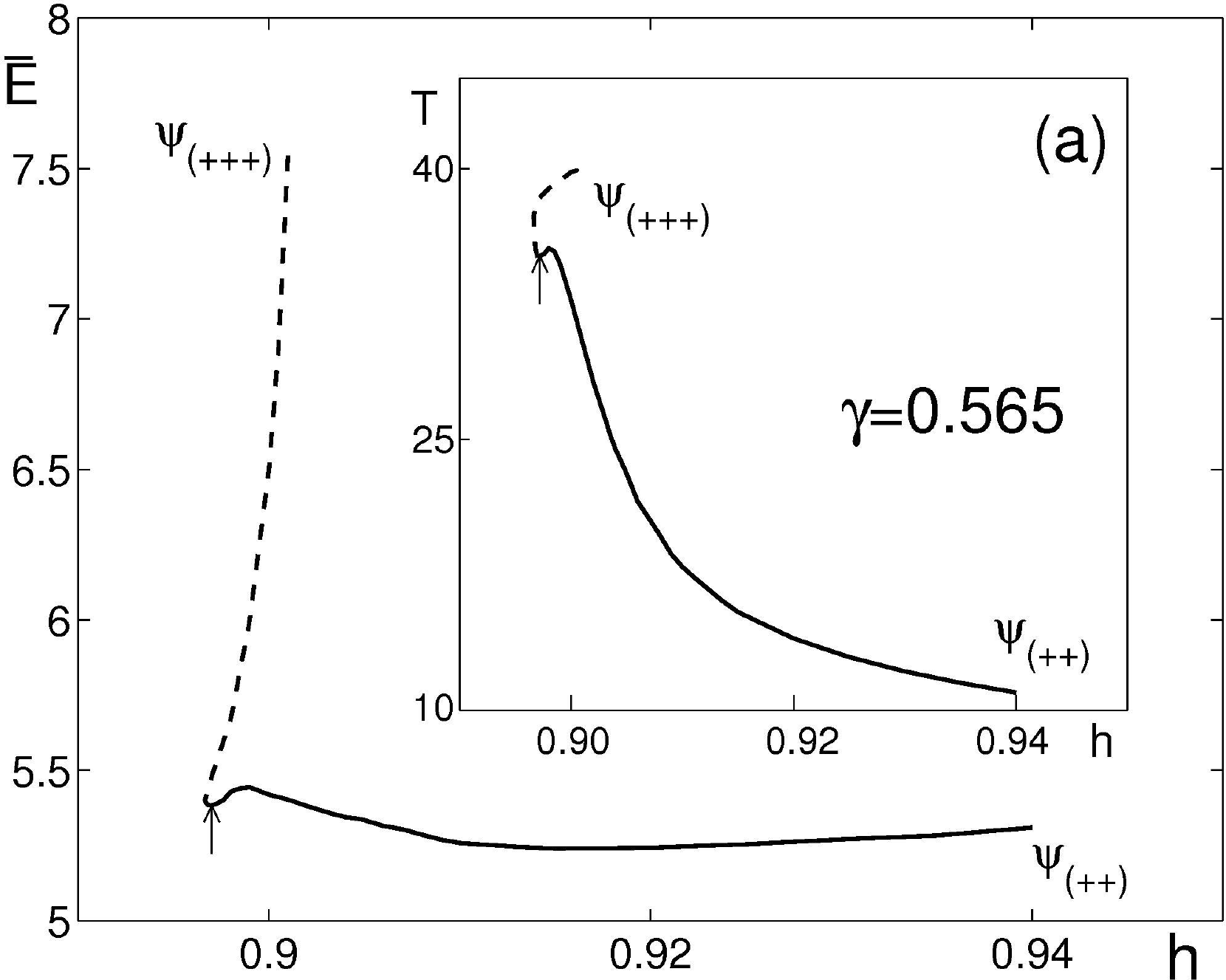}

\vspace*{1mm}
\includegraphics[width = \linewidth]{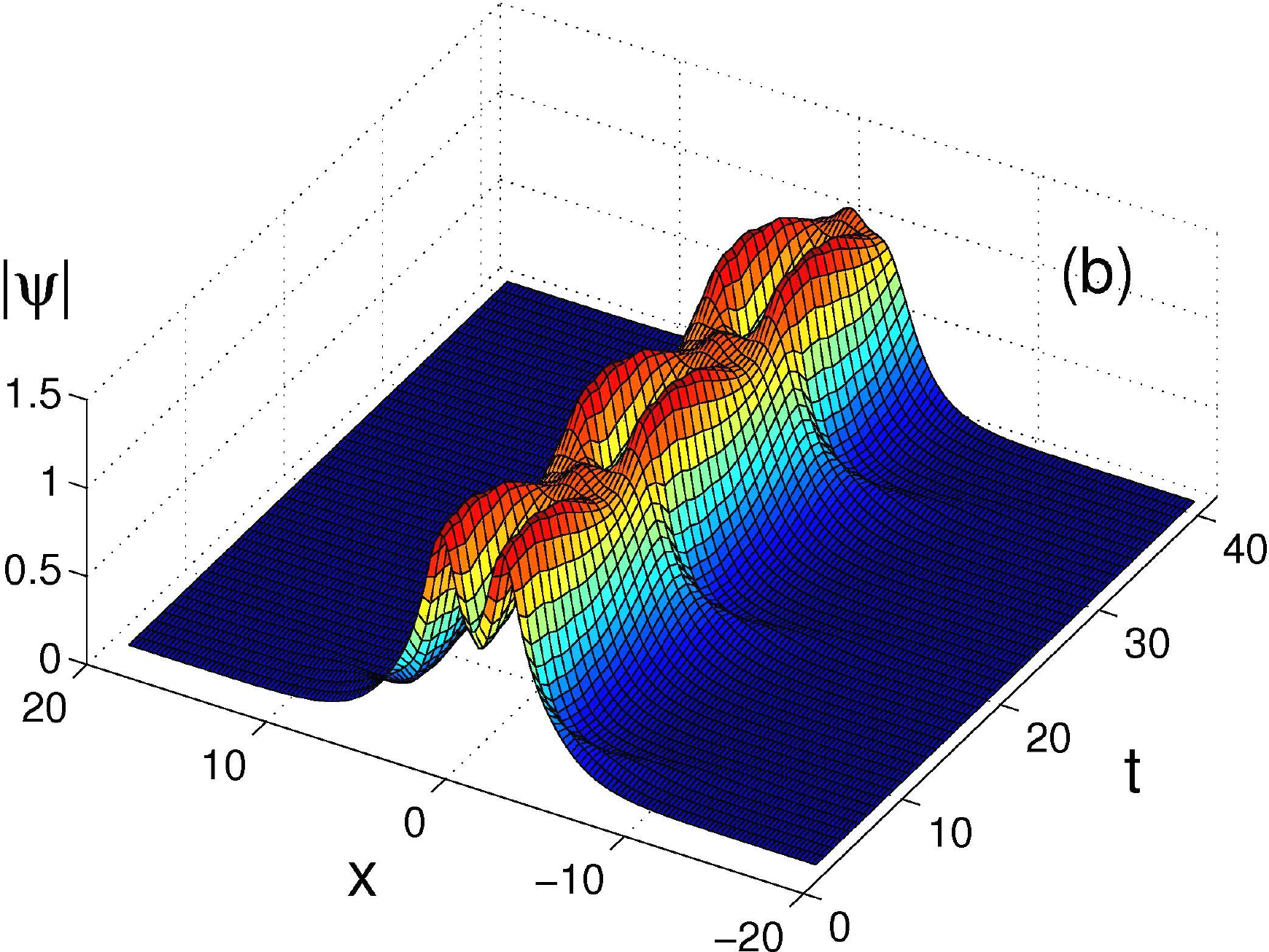}

\vspace*{1mm}
\includegraphics[height = 1.7in, width = \linewidth]{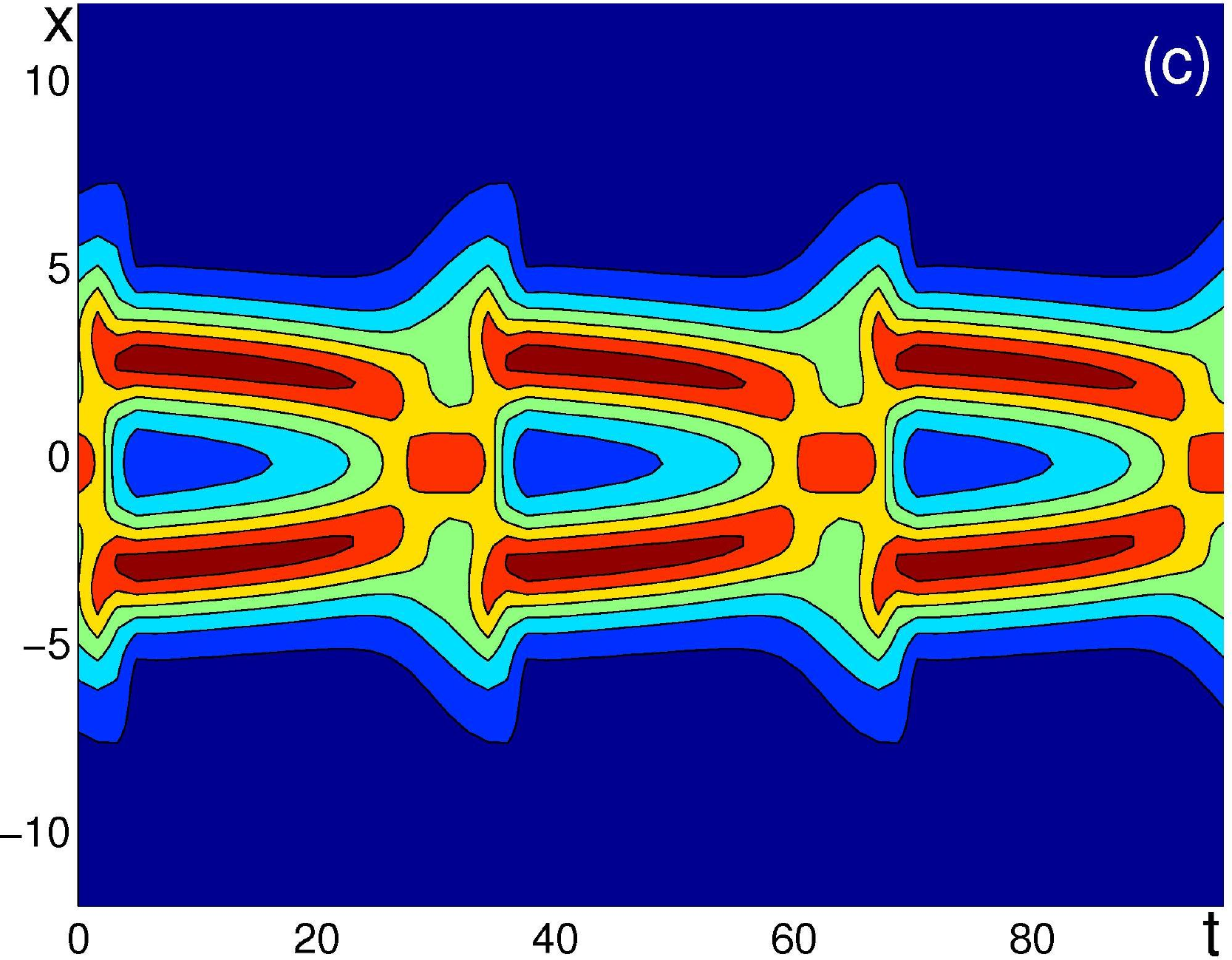}
\caption{\label{565}  (Color online) 
 (a) The average energy (main panel) and the period (inset) of the
periodic solution arising for $\gamma=0.565$. The solid curve shows the
stable and the dashed one unstable branch. The 
point of the period-doubling bifurcation is marked 
 by a vertical arrow. 
(b-c)  Representative solutions on the lower branch in panel (a).
(b): $h=0.92$, $T=13.973$; shown is the absolute value
 of $\psi$. (c): $h=0.90$,  $T=32.729$; shown are the level curves of $|\psi|$.
 The time interval covered by (b) and (c) includes three periods of oscillation.
}
\end{figure}

When $\gamma=0.565$, the  (only) Hopf bifurcation is 
at  $h_{H1}= 0.94$. Using  this value as a starting point in our
continuation process, results in the 
 bifurcation diagram
shown in
Fig.\ref{565}(a).
In order to articulate details  of the diagram, 
  we supplement 
the graph of the period $T(h)$ with a plot of
the averaged energy, defined by
\be {\overline E}= \frac{1}{T} \int_0^T E(t) \, dt, 
\label{Ebar} \ee
where $E(t)$ is given by Eq.\eqref{E}.  
We also evaluate Floquet multipliers as described in \cite{BZvH}.

At the starting point $h_{H1}= 0.94$, the Floquet spectrum includes three unit
eigenvalues and two complex-conjugate pairs with moduli smaller
than one. As $h$ is decreased from $h= 0.94$, two unit
eigenvalues remain in the spectrum while the third one moves
inside the unit circle along the real axis. This positive eigenvalue
decreases  in modulus until it passes to  the negative semiaxis at
$h=0.92$; once the eigenvalue has become negative, it  starts growing in absolute value. Eventually,
as $h$ reaches the value of
$0.897$, the negative real eigenvalue crosses through $\mu=-1$.
A period-doubling bifurcation occurs at this point; as $h$
drops below $0.897$, the periodic complex becomes unstable but a stable
double-periodic solution is born. Note that the destabilization
occurs {\it not\/} at the turning point of the ${\overline E}(h)$ curve
(which is at $h=0.89665$) but for a slightly larger $h$,
i.e. before the turning point is reached. 
Fig.\ref{565}(b) shows a representative solution on the lower,
``horizontal",
branch of the ${\overline E}(h)$ curve.

As for the two complex pairs,  the eigenvalues constituting one of
these grow in absolute value as we  move along the ``horisontal" branch 
 towards smaller $h$. At the same time, the imaginary parts of these eigenvalues
 decrease and the pair converges on the positive real axis --- just before crossing through the unit circle.
 The two real eigenvalues cross through $\mu=1$ almost simultaneously,
 as the curve turns back at $h=0.89665$; after that, 
they remain outside the unit circle.
 The other complex pair also converges on the real axis but
 remains
 inside the unit circle along the entire curve.

As $h$ is decreased and we approach the turning point in Fig.\ref{565}(a), 
the amplitude of oscillations grows and the solution
transforms into a sequence of soliton fusions and fissions.
Two solitons merge into one entity which then breaks into two constituents,
and this process continues periodically; see Fig.\ref{565}(c).

The whole of the ``vertical" branch of the ${\overline E}(h)$ curve is unstable. The
branch ends at the stationary $\psi_{(+++)}$ solution (here
$h=0.901$).  
As we approach the
 endpoint of this branch, the two real (positive) eigenvalues
 with $\mu < 1$ and one of the two eigenvalues with $\mu>1$
 move closer to 1. At the endpoint, the spectrum includes three unit
 eigenvalues and two  real eigenvalues close to $1$. This  corresponds to the 
 spectrum of a stationary three-soliton complex near its Hopf boifurcation.

\begin{figure}[t]
\includegraphics[width = \linewidth]{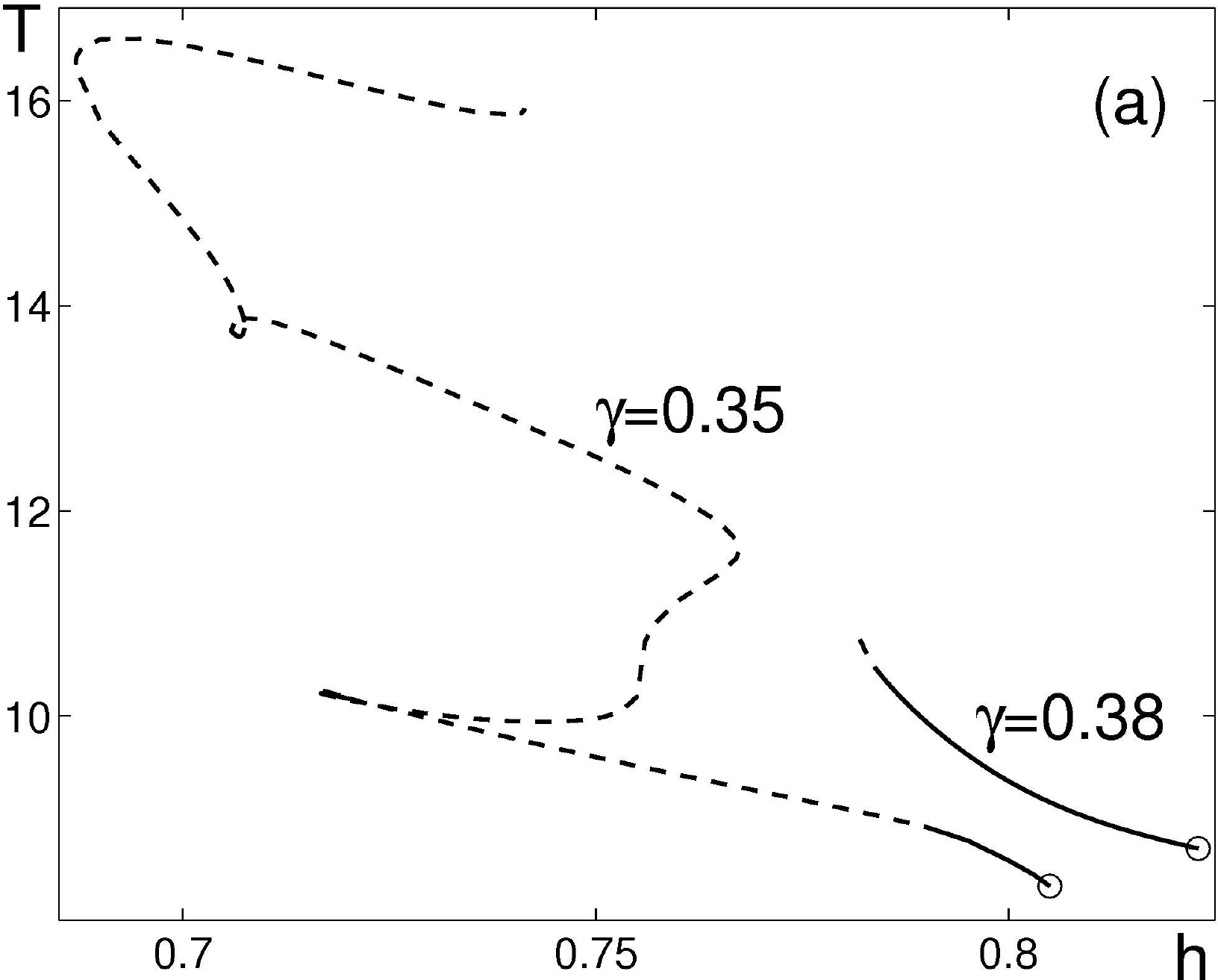}
\includegraphics[width = \linewidth]{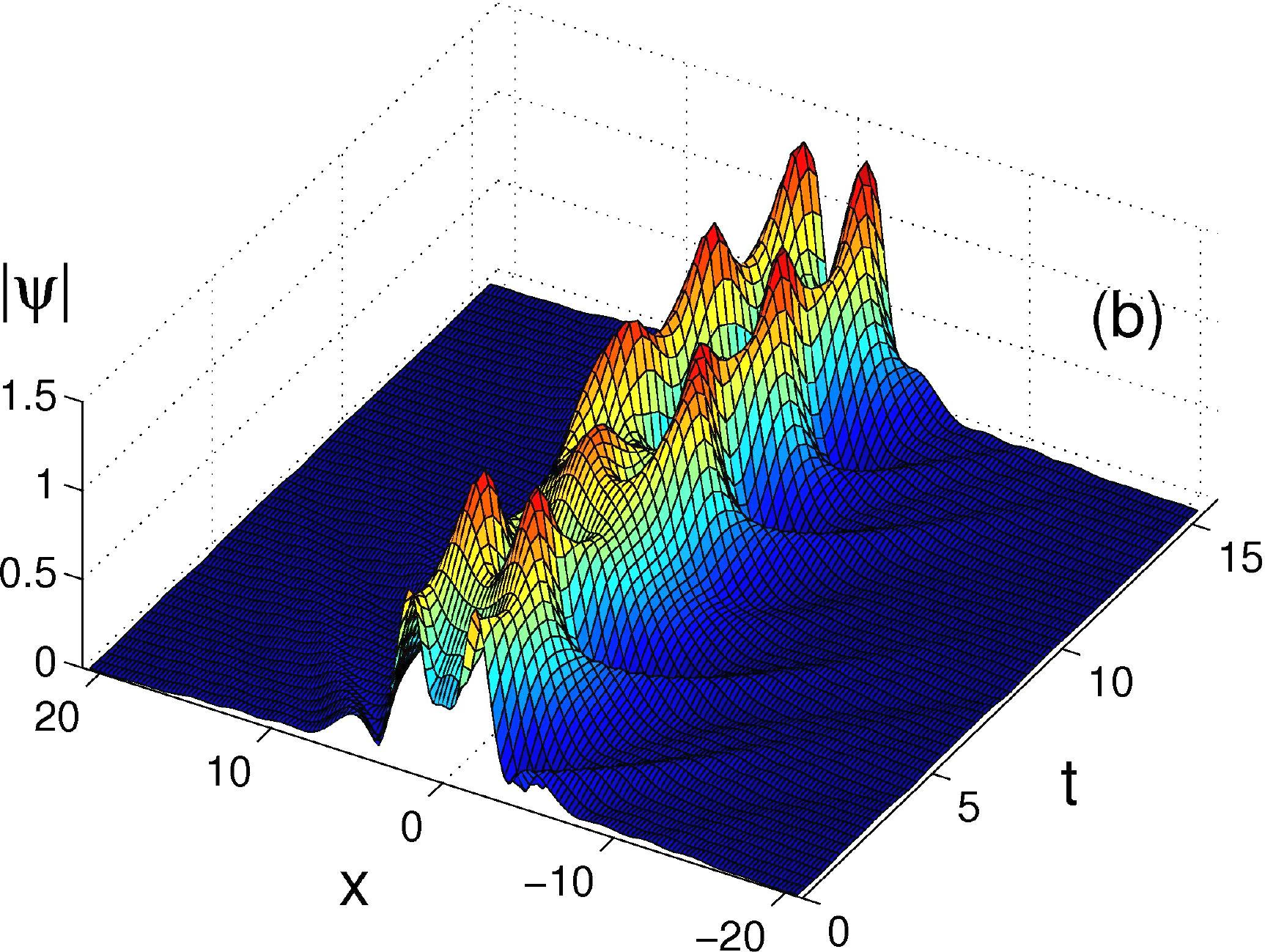}
\caption{\label{pp1} (Color online)
The first branch of the two-soliton
time-periodic complex for $\gamma=0.35$
and $\gamma=0.38$. 
 (a): the period of
the solution.  The solid curves show the
stable and the dashed ones unstable branches.  
The circles mark the starting points of the continuation
(the stationary complex $\psi_{(++)}$). 
(b):  A two-soliton periodic solution with 
complex temporal behaviour  arising
at the end point of the $\gamma=0.35$ curve  in (a).
 (Here $h=0.741$, $T=15.9$.) The figure shows  just one period of oscillation.
}
\end{figure}

Proceeding to the region with more than one Hopf point, we  consider $\gamma=0.35$.
Here, the ``lower" Hopf bifurcation occurs at $h_{H1}=0.805$. This bifurcation
is supercritical; as $h$ drops below $h_{H1}$, 
the stationary two-soliton bound state loses its stability to a periodic two-soliton complex
which is born at this point. At the 
bifurcation point, the spectrum of the Floquet multipliers 
includes three unit eigenvalues and two 
complex-conjugate pairs inside the unit circle  --- one with ${\rm Re\/} \, \mu<0$ and
the other one with ${\rm Re\/} \, \mu >0$.
As we continue the periodic complex 
towards smaller $h$, the negative-real-part pair converges
on the real axis inside the unit circle, after
which  one of the resulting negative
eigenvalues grows in absolute value and, at $h=0.79$, crosses through $\mu=-1$.
The periodic complex loses its stability to a double-periodic bound state of two
solitons.    As we continue the unstable branch, it makes a
number of turns (Fig.\ref{pp1}(a)), 
the spatiotemporal complexity of the solution increases 
(Fig.\ref{pp1}(b)) but it
never regains its stability. 

Another representative value of $\gamma$ with two Hopf bifurcations, is $0.38$.
Here, the continuation of the two-soliton complex from the lower Hopf point results
in the $T(h)$ curve similar to the $\gamma=0.35$ case (Fig.\ref{pp1}(a)). 
Like in the $\gamma=0.35$ case, the solution loses stability in a period-doubling 
bifurcation. We did not path-follow the unstable branch far beyond the bifurcation point.

Summarising results of continuation  from the first, ``lowest",
Hopf bifurcation in Fig.\ref{chart}, we note that the bifurcation is supercritical --- both for large and small $\gamma$. 
Another common feature is the loss of stability resulting from a Floquet multiplier crossing through $\mu=-1$.
Since this bifurcation occurs before the first turn of the $T(h)$ curve, 
it always gives rise to a {\it stable\/} double-periodic solution. 
It is also fitting to note that all time-periodic complexes 
emerging in the first Hopf bifurcation are symmetric in space
(i.e. invariant under the reflection $x \to -x$).

\subsection{The second and the third Hopf bifurcations}

When $\gamma$ lies between $0.34$ and $0.39$, the stationary two-soliton complex
suffers {\it two\/}  Hopf bifurcations, at $h_\textrm{H1}$ and  $h_\textrm{H2}$,
with $h_\textrm{H2}> h_\textrm{H1}$. (These are marked 1 and 2 in Fig.\ref{chart}.)
In this subsection, we describe the continuation of periodic solutions  detaching at $h_\textrm{H2}$ 
(the second of the two bifurcations)
for several representative values of damping.

\begin{figure}[t]
\includegraphics[width = \linewidth]{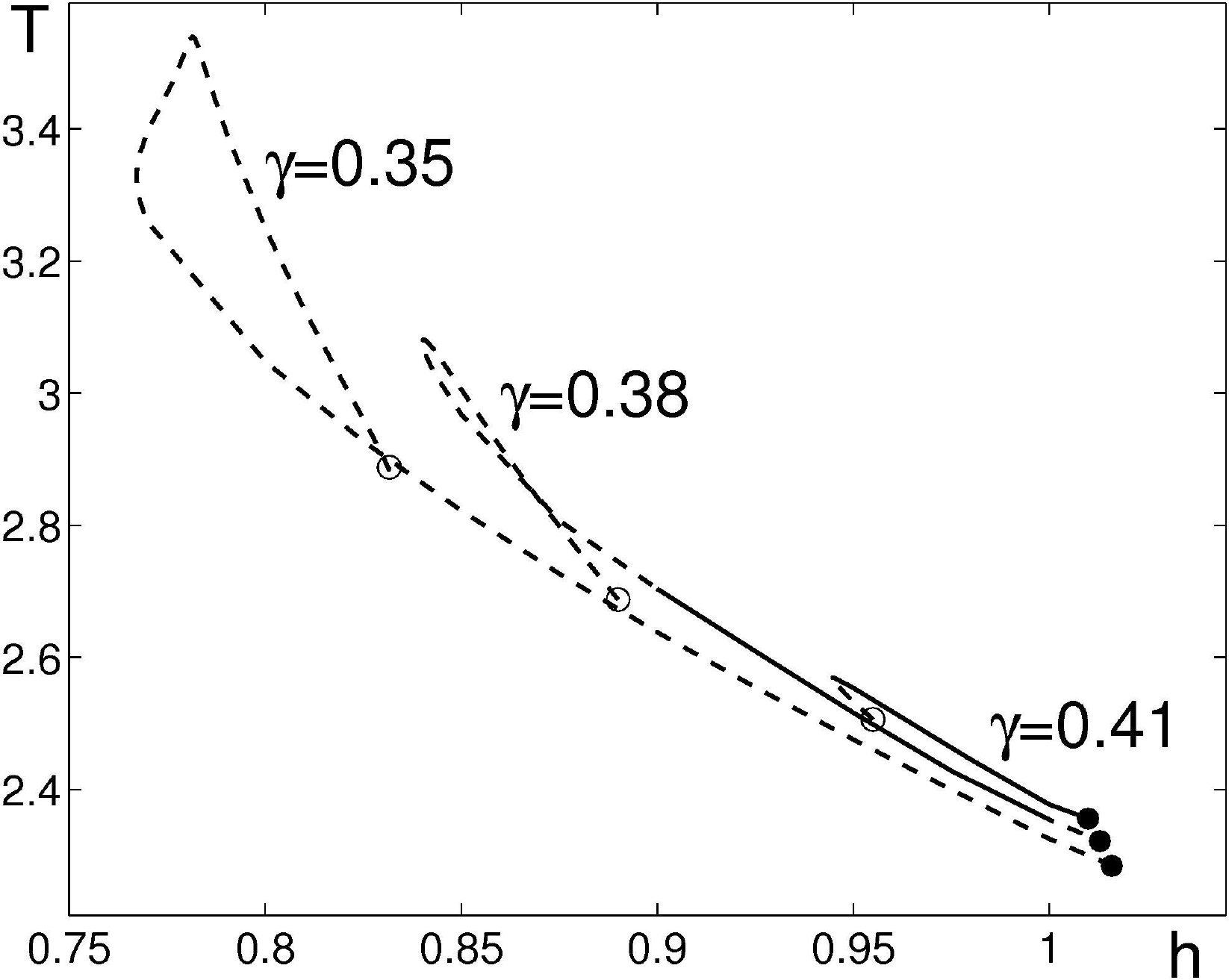}
\caption{
\label{pp2}
The second branch of the periodic two-soliton solution.
The empty circles mark the 
starting points of the
continuation ---
the points $h=h_\textrm{H2}$ where the stationary two-soliton complex 
suffers the second Hopf bifurcation. The full circles
mark the endpoints. The endpoint of the $\gamma=0.41$ curve corresponds to the third Hopf bifurcation
($h_\textrm{H3}$); the endpoints of the $\gamma=0.35$ and $\gamma=0.38$ curves 
lie inside the instability domain of the stationary complex.
The solid curves show the
stable and the dashed ones unstable branches. }
\end{figure}

The second Hopf bifurcation is subcritical: 
the emerging periodic branch is unstable
and coexists with the stable stationary branch. That is, the periodic branch initially
extends {\it down\/} in  $h$, see  the $\gamma=0.35$ 
and $\gamma=0.38$ curves in Fig.\ref{pp2}. 
At some point the branch turns back after which $h$ grows without any further U-turns;
notably, it grows beyond the  interval of the stable stationary bound states (Fig.\ref{pp2}).

The periodic branch ends at an unstable  stationary complex. The endpoint corresponds
to the ``concealed" Hopf bifurcation of the stationary solution
where a pair of complex-conjugate eigenvalues crosses from one half of the complex
plane to the other but the solution remains unstable due
to additional unstable eigenvalues.

When $\gamma$ is set on $0.35$, the whole periodic branch is unstable but 
for $\gamma$ as close as $0.36$ a narrow stability window appears inside it.
As $\gamma$ grows from $0.36$, the stability window expands
--- see the $\gamma=0.38$ curve in Fig.\ref{pp2}
  which features a sizeable stability interval  $h_1<h<h_2$,
  with  $h_1=0.9415$ and   $h_2=1.015$.
 Within this stability window,  the periodic complex has two complex-conjugate pairs
 of Floquet multipliers
 $\mu_4=\mu_3^*$ and  $\mu_6=\mu_5^*$, with $|\mu|<1$
 (in addition to two unit eigenvalues). 
 As $h$ is decreased below $h_1$, the first  pair ($\mu_{3,4}$) moves 
 outside the unit circle, with the second pair remaining inside; 
 when $h$ is raised above $h_2$, the unit circle is crossed by the second pair 
 ($\mu_{5,6}$), with the first pair remaining inside. Thus the stability 
 interval is bounded by the Neimark-Sacker bifurcation on each side.
 This observation suggests that a quasi-periodic two-soliton complex should be 
 born on the crossing of either stability boundary,
 $h_1$ and $h_2$ --- the conclusion confirmed by direct 
numerical simulations of Eq.\eqref{NLS}.
(Quasiperiodic solutions can obviously not be captured by the periodic boundary-value problem;
the direct numerical simulation remains the only feasible way of determining them.)
   
It is worth mentioning here that the periodic two-soliton complexes 
coexist with periodic one-soliton solutions. (For example, for $\gamma=0.35$, the periodic
free-standing soliton exists between $h=0.75$ and $h=1.02$; see Fig.2 in \cite{BZvH}.)
However the one- and two-soliton branches are not    
connected.

When  $\gamma$ is between $0.39$ and $0.413$,  the stationary complex undergoes 
four Hopf bifurcations,
$h_\textrm{H1} < h_\textrm{H2}< h_\textrm{H3} < h_\textrm{H4}$
(marked 1, 2, 3, 4 in Fig.\ref{chart}.).
This is the interval  of $\gamma$ that contains the top ``peninsula" in Fig.\ref{chart}.
Choosing $\gamma=0.41$ as a representative value of damping,
we path-followed the periodic complex which is bifurcating off at the point $h_\textrm{H2}$.
Like in the case $0.34 < \gamma < 0.39$ discussed above, the bifurcation is subcritical:
the emerging periodic branch is unstable and initially extends {\it down\/} in $h$. As
in the previously discussed case, the $T(h)$ curve turns 
(at $h=h_\textrm{tn}=0.9447$) and the entire subsequent continuation 
proceeds in the direction of increasing $h$ (Fig.\ref{pp2}).

To describe
the motion of the Floquet multipliers, it is convenient to start somewhere within the 
``upper" part of the $\gamma=0.41$ branch, e.g. at $h=0.99$. 
At this point, the spectrum of linearisation includes two
pairs of complex-conjugate multipliers $\mu_4=\mu_3^*$,  $\mu_6=\mu_5^*$, both with $|\mu|<1$.
However, in contrast to the previously discussed scenario, neither of these two pairs crosses through
the unit circle as $h$ is inceased or decreased and so
the periodic complex with this $\gamma$ does not experience any Neimark-Sacker bifurcations.
 
 As $h$ is decreased from $0.99$, the multipliers $\mu_{3,4}$ converge on the real axis
 and, at the turning point  $h_\textrm{tn}$, cross through $\mu=1$ (almost simultaneously).
 The other complex pair, $\mu_{5,6}$, remains inside the unit circle.
 Therefore, the turning point corresponds to a saddle-node bifurcation of limit cycles.
 If we, instead, {\it increase\/} $h$ starting at $h=0.99$, it is the   $\mu_{5,6}$  pair
  that converges on the real axis, just before $\mu_5$ becoming equal to one.
 At this point the periodic branch rejoins the branch of stationary complexes;
 this value of $h$ is nothing but $h_\textrm{H3}$,  the point of the third Hopf bifurcation of 
 the stationary two-soliton bound state. At $h=h_\textrm{H3}$, the Floquet spectrum includes three unit eigenvalues,
 a real eigenvalue $\mu_6$ close to (but smaller than) one, and a complex-conjugate pair $\mu_{3,4}$
 inside the unit circle. Thus the periodic complex remains stable in the
 whole range between the turning point $h_\textrm{tn}$ and the point $h_\textrm{H3}$ where it
 rejoins the (stable) stationary branch.

   Since  the saddle-node bifurcation point $h_{\rm tn}$
lies {\it below\/} $h_{\rm H2}$,  there is an interval $h_{\rm tn} \leq h \leq h_{\rm H2}$
where we have bistability between the stationary and time-periodic two-soliton complexes.

In summary, the second Hopf bifurcation 
 is always subcritical; the continuation 
 connects it either to the third (supercritical) Hopf bifurcation, or to a 
 concealed bifurcation of unstable two-soliton complexes.
 All time-periodic solutions arising in these bifurcations are symmetric in space.

\subsection{The fourth, symmetry-breaking,  Hopf bifurcation}

\begin{figure}[t]
\includegraphics[width = \linewidth]{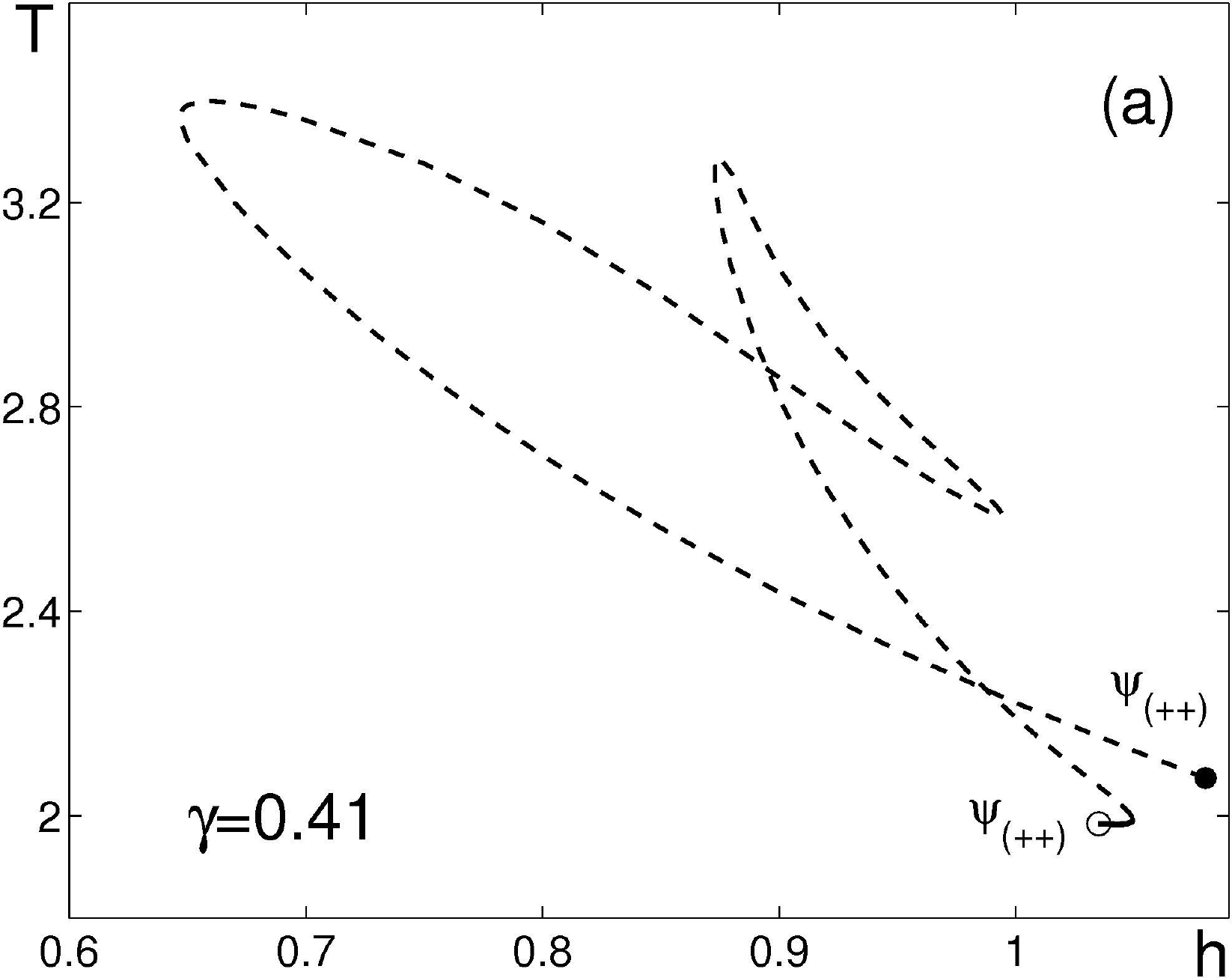}
\includegraphics[width = \linewidth]{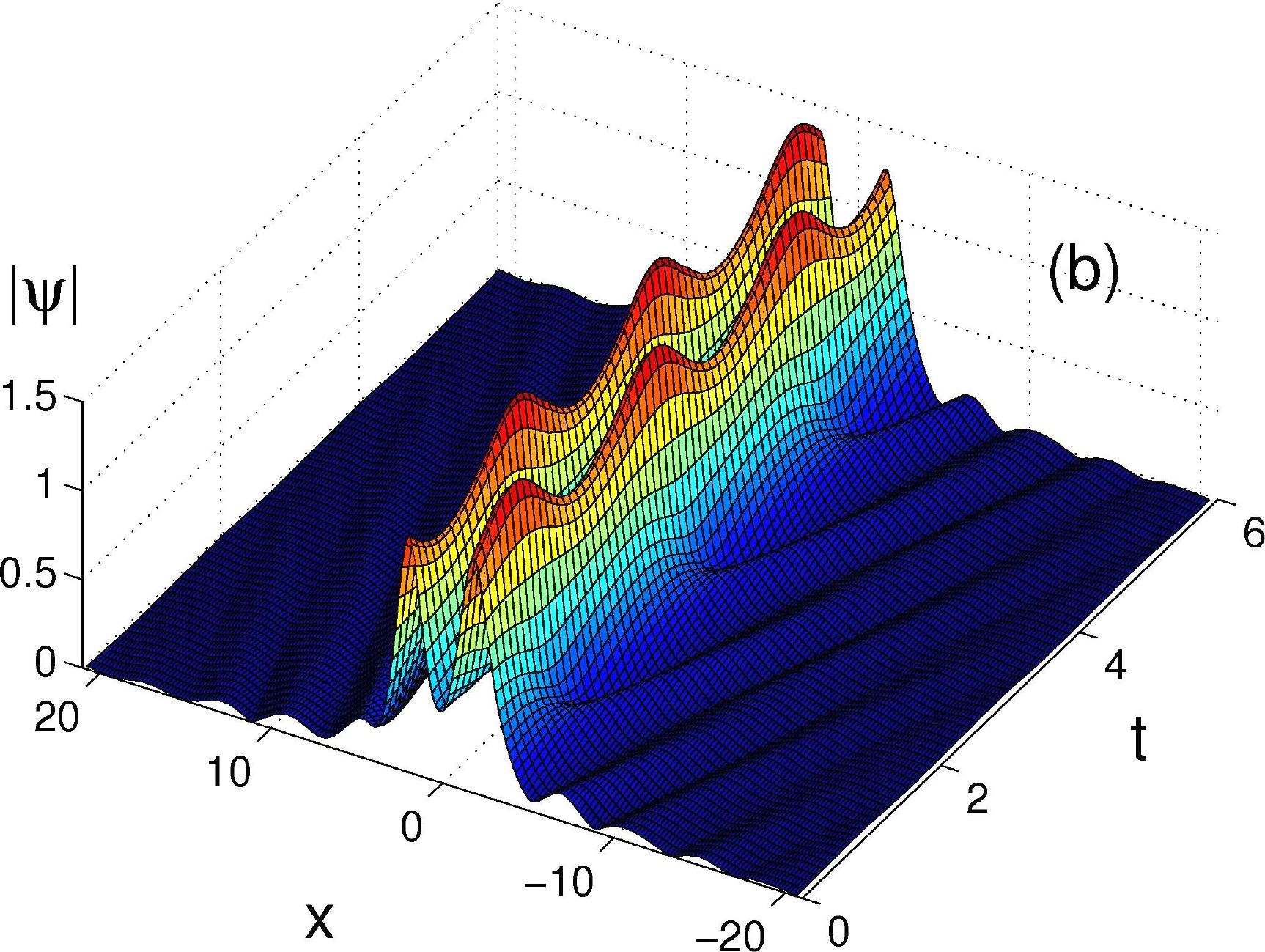}
\caption{\label{asymmetric} (Color online) 
(a) The branch of periodic 2-soliton complexes oscillating out of phase with each other.
 The empty circle marks the 
starting point of the
continuation ---
the point $h=h_\textrm{H4}$ where the stationary two-soliton complex 
suffers the symmetry-breaking Hopf bifurcation.
The full circle marks the endpoint. 
The short solid segment near the beginning of the curve represents the
stable solution;  the rest of the branch (dashed) is unstable.
(b)
A representative  solution on the stable part of this branch. 
Here $h=1.0493$ and $ T=1.991$; several periods of oscillation are shown. }
\end{figure}

The locus of the fourth Hopf bifurcation 
is a stretch of the north-west coast of the 
 ``continent" of stable stationary complexes  in Fig.\ref{chart}
(marked 4).   The ``north-western coastline" extends
 from $\gamma=0.39$ to the point $\gamma=0.445$,
where it  meets the continuous-spectrum instability curve $h=h_\textrm{cont}(\gamma)$.
At the bifurcation point
a pair of complex eigenvalues $\lambda, \lambda^*$ of the
eigenvalue problem
\[
(\mathcal{H}-\gamma J) {\bf p}(x)= \lambda J {\bf p}(x),
\]
cross through the imaginary axis.
Here $\mathcal{H}-\gamma J$
is the operator of  linearisation about the stationary solution (see \cite{BZvH} for details). 
The  bifurcation is symmetry breaking:
unlike three other Hopf bifurcations, the corresponding eigenfunctions 
${\bf p}(x)$ and ${\bf p}^*(x)$ are odd (antisymmetric):
${\bf p}(-x)= -{\bf p}(x)$. 
Accordingly,  the time-periodic solutions 
which are born in this bifurcation describe out-of-phase oscillations
 of 
two identical solitons making up the complex
[see Fig.\ref{asymmetric}(b)].

The fourth Hopf bifurcation is supercritical: 
the emerging periodic branch is stable and extends {\it up\/} in $h$.
For $\gamma=0.41$ which we take as a representative value of damping,
this bifurcation occurs at $h_{\rm H4}=1.037$. 
At the bifurcation point, the Floquet spectrum includes three 
eigenvalues $\mu_{1,2,3}=1$, two real positive multipliers $\mu_{4,5}<1$
and several complex-conjugate pairs with $|\mu|<1$. 
As $h$ grows from $h_\textrm{H4}$, one of the unit eigenvalues moves inside the
unit circle, but as $h$ is further increased, it reverses and moves out. 
At this point ($h=1.049$), a saddle-node bifurcation 
of cycles occurs; the periodic solution loses its stability and the 
branch turns back [Fig.\ref{asymmetric}(a)].
As we continue it further, the real and complex eigenvalues move back and forth through the 
unit circle; some pairs converge on the real axis --- but the solution never regains its stability.

After a lengthy excursion into the $(h,T)$ plane [Fig.\ref{asymmetric}(a)], 
the periodic branch rejoins the branch of (unstable) stationary complexes $\psi_{(++)}$
(at $h_\textrm{cnc}=1.082$). 
The spectrum of the endpoint stationary solution   includes three unit eigenvalues, a positive eigenvalue $\mu_4<1$,
and two complex-conjugate pairs, with $|\mu_{5,6}|<1$ and $|\mu_{7,8}|>1$.
The value $h=h_\textrm{cnc}$ pertains to the concealed Hopf bifurcation of the unstable stationary complex $\psi_{(++)}$.

\section{The two-soliton attractor chart
and open problems}
\label{DC}

\begin{figure}[t]
\includegraphics[width = \linewidth]{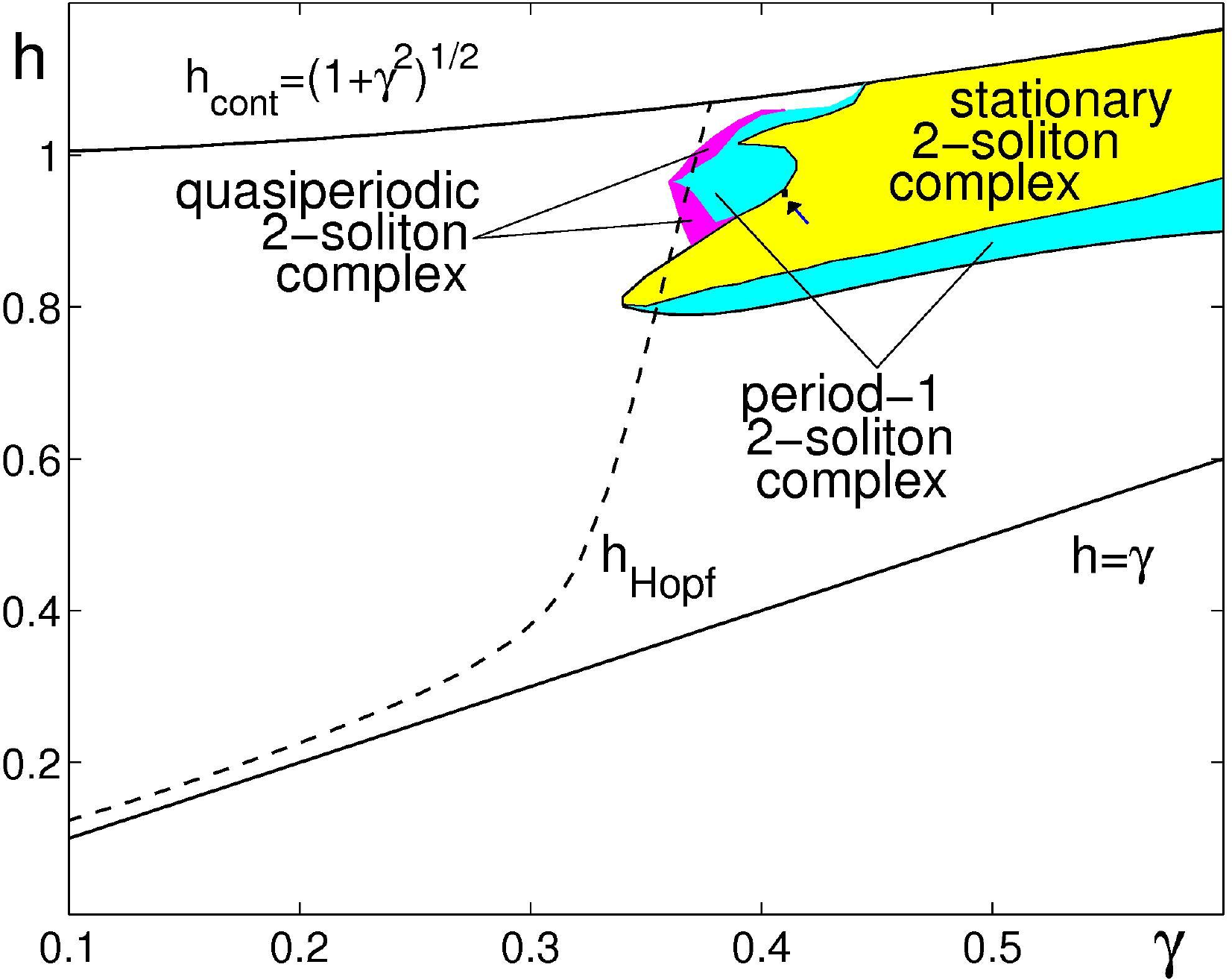}
\caption{\label{chart_b} (Color online) 
 Two-soliton attractor chart. 
The  chart is still under construction;   the outer boundaries of the purple (dark gray)
area (populated by 
quasiperiodic two-soliton attractors)
are still to be refined. The black mark 
(pointed to by the arrow) initiates the periodic/stationary 
bistability region.
The dashed curve is the line of the  
Hopf bifurcation of the $\psi_+$ soliton, shown just for the reference purposes.}
\end{figure}

Fig.\ref{chart_b} summarises our results on the stationary and  periodic two-soliton 
attractors.  This diagram complements the
 single-soliton attractor chart compiled in the first part of this project \cite{BZvH}. 
The two-soliton chart  is in qualitative agreement with results of  direct numerical
simulations \cite{Wang}. 

In Fig.\ref{chart_b},
 we have included stable quasiperiodic complexes  [highlighted in purple (dark gray)].
 The boundaries between the stable-periodic and stable-quasiperiodic domains 
 are defined by the Neimark-Sarker bifurcations of the periodic complexes;
 these admit an accurate demarcation using our method (i.e. by monitoring the Floquet multipliers
 along the periodic branches). On the other hand,
 in order to determine where the stable quasiperiodic solution ceases to exist, 
 we had to relinquish our 
continuation approach in favour of
direct numerical simulations of Eq.\eqref{NLS}. 
We have performed only a few runs
and hence Fig.\ref{chart_b} gives only a schematic position of the outer boundary of the 
quasiperiodic stability domain. In order to demarcate this boundary more accurately, 
one would need to 
 perform  numerical simulations more extensively. This is beyond
 the scope of our present study. 

The region of bistability of stationary and periodic complexes also
needs to be accurately
delimited.  So far, we have only demarcated a small portion of it;
see the black mark in Fig.\ref{chart_b}.

Finally, it would also be interesting to continue  periodic solutions bifurcating from the
stationary complexes in the
``concealed" Hopf bifurcations, where the stationary solution 
remains unstable on both sides of the bifurcation due to additional eigenvalues with
positive real parts. 
In our continuation process bifurcations of this sort would typically arise 
as the endpoints of the periodic branches starting at the proper Hopf bifurcations
of the stationary complexes. Starting at the concealed bifurcations would 
produce an additional wealth of periodic branches some of which may have stable segments.

\acknowledgments

We thank Nora Alexeeva for providing us with the 
simulation code for 
equation \eqref{NLS}. An instructive conversation with Alexander Loskutov
is gratefully acknowledged.
IB was supported by the NRF of South Africa
(grants UID 65498, 68536 and 73608).
 EZ was supported by a DST grant under the
JINR/RSA Research Collaboration Programme and partially supported
by RFBR (grant No. 09-01-00770).


\begin{thebibliography}{99}


\bibitem{BZvH} I V Barashenkov, E V Zemlyanaya, T C van Heerden,
previous submission 

\bibitem{MEMS}
E. Kenig. B. A. Malomed, M.C. Cross, and R. Lifshitz, Phys. Rev. E {\bf 80}, 046202 (2009);
M. Syafwan, H. Susanto, and S. M. Cox, Phys. Rev. E {\bf 81}, 026207 (2010)

\bibitem{DCF}
N. Dror and B. A. Malomed, Phys. Rev. E {\bf 79},  016605 (2009)



\bibitem{OPO}
K. Staliunas, J. Mod. Optics {\bf 42}, 1261 (1995);
S. Longhi, Optics Lett. {\bf 20}, 695 (1995);
S. Longhi and A. Geraci, Appl. Phys. Lett. {\bf 67}, 3060 (1995);
S. Longhi. Phys. Scr. {\bf 56}, 611 (1997);  
S. Longhi, G. Steinmeyer, and W. S. Wong, J. Opt. Soc. A. B {\bf 14}, 2167 (1997);
K. Promislow and J N Kutz, Nonlinearity {\bf 13}, 675 (2000); 
R. O. Moore, K. Promislow, Physica D {\bf 206}, 62 (2005);
I.  P\'erez-Arjona, E. Rold\'an, and G. J. de Valc\'arcel, Phys. Rev. A {\bf 75}, 063802 (2007)




\bibitem{DB} 
D. Hennig, Phys. Rev. E {\bf 59}, 1637 (1999);
Y. Feng, W.-X. Qin, Z. Zheng, Phys. Lett. A {\bf 346}, 99 (2005);
H. Susanto, Q. E. Hoq, and P. G. Kevrekidis, Phys. Rev. E {\bf 74}, 067601 (2006);
J. Garnier, F. Kh. Abdullaev, and M. Salerno, Phys. Rev. E {\bf 75}, 016615 (2007)

\bibitem{comp}
  S. Wabnitz, Opt. Lett. {\bf 18}, 601 (1993);
    B. A. Malomed, Phys. Rev. E {\bf 47}, 2874 (1993);
   D. Cai, A.R. Bishop, N. Gr¿nbech-Jensen, and B.A. Malomed,
 Phys. Rev. E {\bf 49} 1677 (1994);
 S. Longhi, Phys. Rev. E {\bf 53}, 5520 (1996);  Phys. Rev. E {\bf 55} 1060 (1997);
   N.N. Akhmediev, A. Ankiewicz, and J.M. Soto-Crespo,
Phys. Rev. Lett. {\bf 79}, 4047 (1997);        
M. Bogdan and A. Kosevich, Proc. Estonian Acad. Sci. Phys. Math. {\bf 46}, 14 (1997);
B. Sandstede,  C. K. R. T.  Jones, J. C.  Alexander,  Physica D {\bf 106}, 167 (1997);
B. Sandstede, Trans. Amer. Math. Soc. {\bf 350}, 429 (1998);
Yu S  Kivshar, A R  Champneys, D Cai, A R Bishop, Phys Rev B {\bf 58}, 5423 (1998);
    V S Gerdjikov, E G Evstatiev, D J Kaup, G L Diankov, I M Uzunov, Phys. Lett. A {\bf 241}, 323 (1998);
         M. Kollmann, H.W. Capel, and T. Bountis,  Phys. Rev. E {\bf 60}  1195 (1999);     
         J. Christoph, M Eiswirth, N Hartmann, R Imbihl, I Kevrekidis, M B\"ar,
         Phys. Rev. Lett. {\bf 82}, 1586 (1999);
M. Or-Guil, I G Kevrekidis, M B\"ar,  Physica D  {\bf 135}, 154 (2000);
M. M. Bogdan, A. M. Kosevich, G. A. Maugin, Wave Motion {\bf 34}, 1 (2001);
T Kapitula, P G Kevrekidis, B A Malomed, Phys Rev E {\bf 63}, 036604 (2001); 
I V Barashenkov, S R Woodford and E V Zemlyanaya, Phys Rev Lett {\bf 90},  054103 (2003);
 I V Barashenkov and S R Woodford, Phys Rev E {\bf 71},  026613 (2005);
O V Charkina, M. M. Bogdan. Symmetry, Integrability and Geometry: Methods and Applications {\bf 2},  047 (2006);
J. M. Soto-Crespo, Ph. Grelu, N. Akhmediev and N. Devine, Phys. Rev. E {\bf 75}, 016613 (2007);
I V Barashenkov, S R Woodford and E V Zemlyanaya, Phys Rev E  {\bf 75},  026604  (2007);
Prilepsky J. E., Derevyanko S. A., Turitsyn S. K.,  J. Opt. Soc. Am. B {\bf 24}, 1254 (2007);
D. Turaev, A. G. Vladimirov, and S. Zelik, Phys. Rev. E {\bf 75}, 045601 (2007); 
A. Zavyalov, R. Iliew, O. Egorov, F. Lederer, Opt. Lett. {\bf 34}, 3827 (2009);
Y. Fang and J. Zhou, J. Russ. Laser Research {\bf 30}, 260 (2009);
K. Zhou, Z. Guo, and S. Liu, J. Opt. Soc. Am. B {\bf 27}, 1099 (2010);
B. Orta\c{c},  A. Zaviyalov, C. K. Nielsen, O. Egorov, R. Iliew, J. Limpert, F. Lederer, A. T\"unnermann, 
Opt. Lett. {\bf 35}, 1578 (2010)

 
 
\bibitem{Mal_Par}
B. A. Malomed, Phys. Rev. E {\bf 47}, 2874 (1993)
 
\bibitem{tails}
K. A. Gorshkov, L. A. Ostrovsky, Physica D {\bf 3}, 428 (1981);  
T. Kawahara, S. Toh, Phys. Fluids {\bf 31},  2103 (1988); 
B.A. Malomed,  Phys. Rev. A  {\bf 44}  6954 (1991);
A V Buryak, N N Akhmediev, Phys Rev E {\bf 51}, 3672 (1995);
C. I. Christov, G. A. Maugin, M. G. Velarde, Phys. Rev. E {\bf 54}, 3621 (1996);
I. V. Barashenkov, Yu. S. Smirnov, N. V. Alexeeva, Phys. Rev. E {\bf 57}, 2350 (1998) 
W. Chang, N.  Akhmediev,  and S. Wabnitz, Phys. Rev. A {\bf 80}, 013815 (2009)

\bibitem{embed} 
A V Buryak, Phys Rev E {\bf 52}, 1156 (1995); 
D C Calvo, T R Akylas,  Physica D {\bf 101}, 270 (1997); 
J Fujioka, A Espinosz, J Phys Soc Jpn {\bf 66}, 2601 (1997); 
A R Champneys, B A Malomed, M J Friedman, Phys Rev Lett {\bf 80}, 4168 (1998);
A. R. Champneys, Yu. S. Kivshar, Phys. Rev. E {\bf 61}, 2551 (2000);  
K. Kolossowski,  A. R. Champneys, A. V. Buryak, R. A. Sammut, Physica D {\bf 171}, 153 (2002)

\bibitem{BZ1} I V Barashenkov and E V Zemlyanaya,
Phys Rev Lett {\bf 83} 2568 (1999)


\bibitem{water}
J. Wu, R. Keolian, and I. Rudnick, Phys. Rev. Lett. {\bf 52}, 1421 (1984);
 X. Wang and R. Wei, Phys. Lett. A {\bf 192}, 1 (1994); 
 W. Wang, X. Wang, J. Wang, and R. Wei, Phys. Lett. A {\bf 219}, 74 (1996); 
 X. Wang and R. Wei, Phys. Rev. Lett. {\bf 78}, 2744 (1997); 
M. G. Clerc, S. Coulibaly, N. Mujica, R. Navarro, and T. Sauma, Phil. Trans. R. Soc. A {\bf 367}, 3213 (2009) 




\bibitem{BW} I V Barashenkov, S R Woodford, Phys. Rev. E {\bf 75} 026605 (2007)

\bibitem{long_co}
X. Wang and R. Wei,  Phys. Rev. E {\bf 57}, 2405 (1998)

%
\bibitem{Baer} I V Barashenkov, E V Zemlyanaya, and M B\"ar,
Phys Rev E {\bf 64} 016603 (2001);

\bibitem{BZ2}
I V Barashenkov and E V Zemlyanaya,
SIAM J Appl Math  {\bf 64},  800  (2004);
R. O. Moore. Travelling waves in thermally driven optical parametric oscillators.
Talk at the SIAM Conference on Nonlinear Waves and Coherent Structures
(August 2010, Philadelphia, PA)%

\bibitem{Wang} X.Wang, Phys. Rev. E {\bf 58}, 7899 (1998)

%
%
%
%




\end{thebibliography}
\end{document}